\documentclass[aps,prl,twocolumn,showpacs,floatfix,superscriptaddress]{revtex4}
\usepackage{graphicx}

\begin{document}
\flushbottom

\title{
Very Low Temperature Tunnelling Spectroscopy in the heavy fermion
superconductor PrOs$_4$Sb$_{12}$}
\author{H. Suderow}
\affiliation{Laboratorio de Bajas Temperaturas, Departamento de F\'isica de
la Materia Condensada \\ Instituto de Ciencia de Materiales
Nicol\'as Cabrera, Facultad de Ciencias \\ Universidad
Aut\'onoma de Madrid, 28049 Madrid, Spain}
\author{S. Vieira}
\affiliation{Laboratorio de Bajas Temperaturas, Departamento de F\'isica de
la Materia Condensada \\ Instituto de Ciencia de Materiales
Nicol\'as Cabrera, Facultad de Ciencias \\ Universidad
Aut\'onoma de Madrid, 28049 Madrid, Spain}
\author{J. D. Strand}
\affiliation{Ames Laboratory and Departament of Physics and Astronomy \\
Iowa State University, Ames, Iowa 50011, USA}
\author{S. Bud'ko}
\affiliation{Ames Laboratory and Departament of Physics and Astronomy \\
Iowa State University, Ames, Iowa 50011, USA}
\author{P. C. Canfield}
\affiliation{Ames Laboratory and Departament of Physics and Astronomy \\
Iowa State University, Ames, Iowa 50011, USA}
\date{\today}

\begin{abstract}
We present scanning tunnelling spectroscopy measurements on the
heavy fermion superconductor PrOs$_4$Sb$_{12}$. Our results show
that the superconducting gap opens over a large part of the Fermi
surface. The deviations from isotropic BCS s-wave behavior are
discussed in terms of a finite distribution of values of the
superconducting gap.
\end{abstract}

\maketitle

The intriguing magnetic and superconducting properties of most heavy fermion metals,
which include e.g. coexistence of superconductivity with ferromagnetism,
non-Fermi liquid behaviors or multiple superconducting phases,
represent a challenge
to our current understanding of condensed matter physics\cite{Hewson,Aoki01,Hasselbach89}.
 The route towards the discovery of new heavy fermions is
to synthesize materials that maintain degrees of freedom (e.g. local magnetic or electric
moments), which do not undergo a phase transition upon cooling and are coupled to the electron bath.
A large entropy can then be preserved down to low temperatures, and transferred
 over to the electrons.
Pr$^{3+}$ ions offer this possibility if the Pr is in a cubic
point symmetry, where the crystal electric field ground state can
be a non-magnetic, non-Kramers $\Gamma_3$ doublet, which can
provide the entropy needed to create a heavy fermion ground state
if a cooperative Jahn-Teller transition can be
avoided\cite{Yatskar96,Cox87}. PrInAg$_2$ meets these criteria, and it
was shown to be the first possible example of a Pr-based heavy
fermion with a very large linear specific heat coefficient of 6.5
J/mol K$^2$\cite{Yatskar96}. Unfortunately PrInAg$_2$ did not
superconduct down to 50mK, making it hard to independently
ascertain that the large linear specific heat term was indeed of
electronic origin.  Recently superconductivity has been found with
T$_c$=1.85K in another Pr based heavy fermion\cite{Bauer02},
PrOs$_4$Sb$_{12}$, where large electronic specific heat
coefficients $\gamma$ between 300 and 500 mJ/mol K$^2$ have been
reported\cite{Bauer02,Vollmer03,MapleLa,Measson,Aoki02}.
 The height of the jump of the specific heat at the superconducting transition,
 which compares well with BCS theory,
 shows that the large $\gamma$ is indeed associated to the conduction electrons,
 and that, in addition, the heavy electronic bath superconducts
 \cite{Bauer02,Vollmer03,MapleLa,Measson}.
Moreover, two superconducting transitions have been clearly
resolved in the specific heat at 1.6 and 1.85 K in high quality
single crystalline
 samples, giving strong indications for the presence of multiple superconducting
 phases\cite{Vollmer03,MapleLa,Measson}.
 The situation seems analogous to the only other known stochiometric
superconductor that presents multiple superconducting phases,
UPt$_3$\cite{Hasselbach89,Joynt02}.
In that case, most present theoretical and experimental scenarios associate
 the Cooper pairing mechanism that leads to multiple phase superconductivity
to magnetic fluctuations\cite{Joynt02}. Instead, in
PrOs$_4$Sb$_{12}$, interactions with fluctuating quadrupolar
(electric) moments seem the most likely mechanism that drives the
system to an unconventional, multiple phase superconducting
state\cite{Bauer02,Vollmer03,MapleLa,Measson,Aoki02,
Kotegawa03,MacLaughlin02,Izawa02}. Note also
 that the isostructural compound LaOs$_4$Sb$_{12}$, which
does not show heavy fermion behavior\cite{Sugawara02},
superconducts with a lower critical temperature (1K
\cite{MapleLa}).

 The determination of the most fundamental superconducting properties
of PrOs$_4$Sb$_{12}$ is clearly needed to understand the
formation of unconventional, multiple phase superconductivity.
 One of the first and most
important points is to try to resolve the structure of the
superconducting gap in the low temperature, low magnetic field
phase, which occupies the largest part of the phase diagram\cite{Vollmer03}.
 Indirect information can be obtained by
thermodynamic, transport, magnetic or NMR measurements, but the
experiments that have been done up to now lead to contradictory
results, and are therefore not conclusive. Specific heat does not
seem adequate to study the superconducting gap, because it shows a
high Schottky peak at low temperatures\cite{Bauer02,Vollmer03,MapleLa,Measson,Aoki02}.
 NQR measurements show
an exponential decrease of 1/T$_1$T at low
temperatures\cite{Kotegawa03} associated to a well developed gap.
The London penetration depth, as measured with muon spin relaxation,
also appears to decrease exponentially at low
temperatures\cite{MacLaughlin02}. On the other hand, the angular
dependent thermal conductivity under magnetic fields appears to be
strongly modulated due to significant changes in the
superconducting gap over the Fermi surface. This result has been
associated to the presence of point nodes\cite{Izawa02}.

Here we present direct measurements of the superconducting gap in
PrOs$_4$Sb$_{12}$, done with high resolution tunnelling
spectroscopy studies in the superconducting phase with a Scanning Tunneling Microscope (STM). We
find a superconducting density of states with no low energy
excitations and a well developed superconducting gap.

We use a home built STM unit and electronics installed in a
partially home built dilution refrigerator. We have previously
tested our experimental set-up by measuring the superconducting
properties of Al, which is possibly the best known superconducting
material with a critical temperature (1.2K) of the same order of
magnitude as PrOs$_4$Sb$_{12}$ (1.85K). The sharpest
superconducting features that we could resolve up to now are the
quasiparticle peaks shown in Fig.1a, where we represent the
tunnelling spectroscopy when both tip and sample are of Al
(prepared following \cite{Suderow02a}).  The width of these
features can be taken as a measure of the resolution in energy of
our spectrometer, and therefore of the effective lowest
temperature of our measurements. We find a width at half maximum
of 16$\mu$V, which corresponds to 184mK. On the other hand, the
best fit to the tunnelling spectra taken between a normal tip of
Au and a sample of Al (Fig.1b) using conventional isotropic BCS
s-wave theory gives the same value for the temperature T=190mK
(within the 5\% uncertainty of the fit; we also use
 $\Delta=175\mu eV$). Therefore, we can take 190mK as the lowest
measuring temperature that can be achieved at present with our
set-up. Our resolution is similar to the one achieved in many
planar junction experiments\cite{Wolf}, with the important
difference that in STM measurements the tunnelling current is
between 4 to 6 orders of magnitude smaller.

\begin{figure}[ht]
\includegraphics[width=8cm,clip]{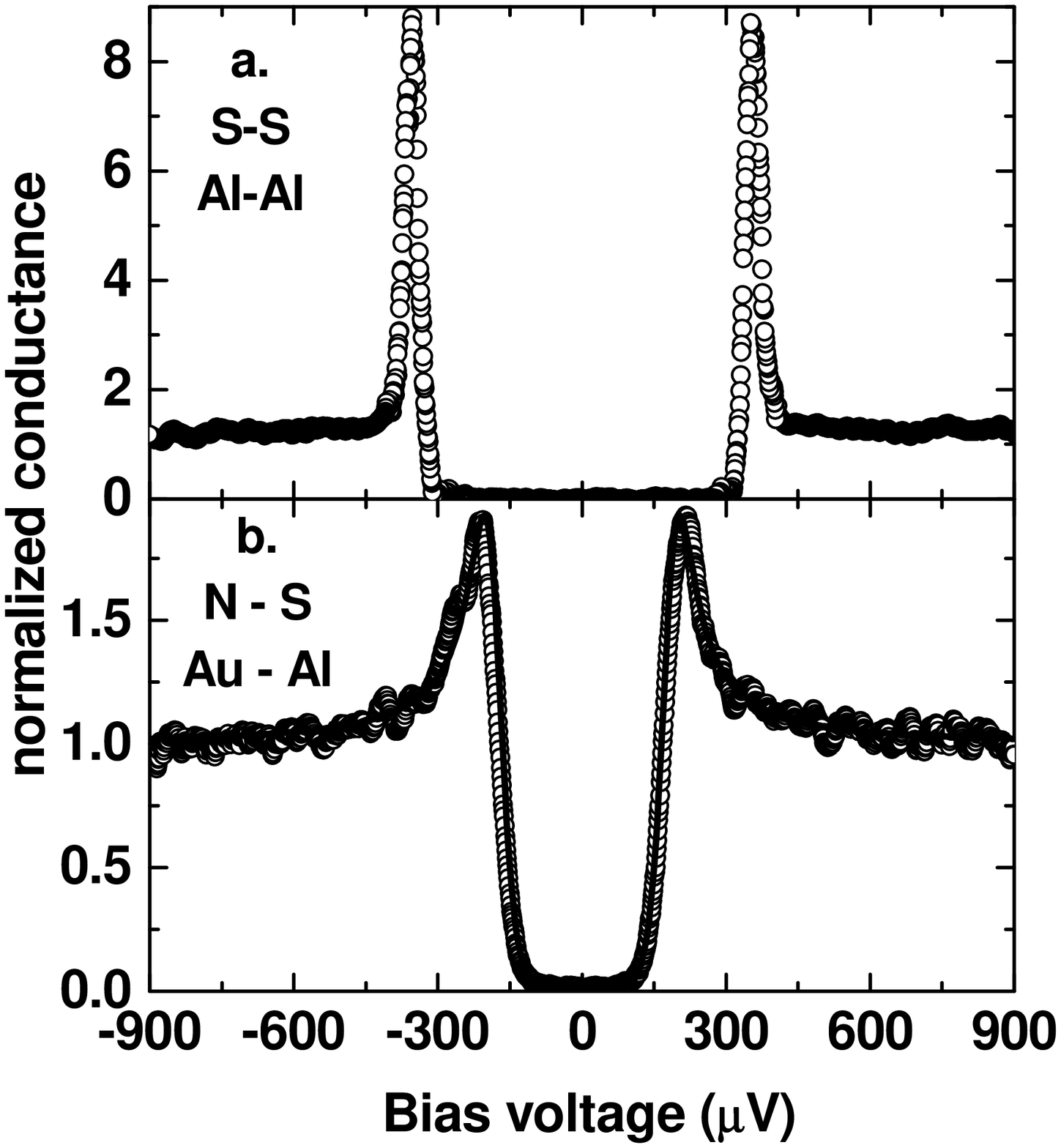}
\vskip -3cm
 \caption{In a. we show the tunnelling
conductance measured at the lowest temperatures of our set-up,
normalized to its value at high voltages (the tunnelling
resistance in this and all subsequent figures is 1M$\Omega$)
 when both tip and sample are of Al. The preparation
procedure is the same as in \protect\cite{Suderow02a}. The peak is
the sharpest superconducting feature that we could resolve with
our experiment and it gives a measure of the resolution of our
spectrometer (184 mK, see text). The tunnelling conductance between an Au tip and a
sample of Al (b.), taken also at the lowest temperatures, gives an
 excellent fit with isotropic, BCS s-wave theory if we take
T=0.19K and $\Delta$=175$\mu$V (line in b.).} \label{fig:Fig1}
\end{figure}

Single crystals of PrOs$_4$Sb$_{12}$ were grown out of a ternary
melt that was rich in both Os and Sb.  A starting composition of
Pr$_2$Os$_{16}$Sb$_{82}$ was heated to 1200 $^\circ$C and cooled
over 100 hours to 725 $^\circ$C and then decanted, revealing small
cubic crystals. We measured three samples in eight different cool
downs by placing an Au tip on optically neat and flat faces of the
single crystals. Within a given cool down, we changed the
macroscopic position of the Au tip on the surface and measured many
(more than 50 in total) different scanning windows
(100x100nm$^2$), using the positioning capabilities of our xy
table, as in previous work \cite{Rubio01}. The measured work
functions were always of several eV and the topography was
reproducible upon changes of the tunnelling current. It consists
of inclined planes and bumps, with typical corrugations of about
4nm (inset of Fig.2), indicating that the crystallographic
direction of the surface is not well defined at the nanoscopic
length scales relevant for this experiment.

\begin{figure}[ht]
\includegraphics[width=8cm,clip]{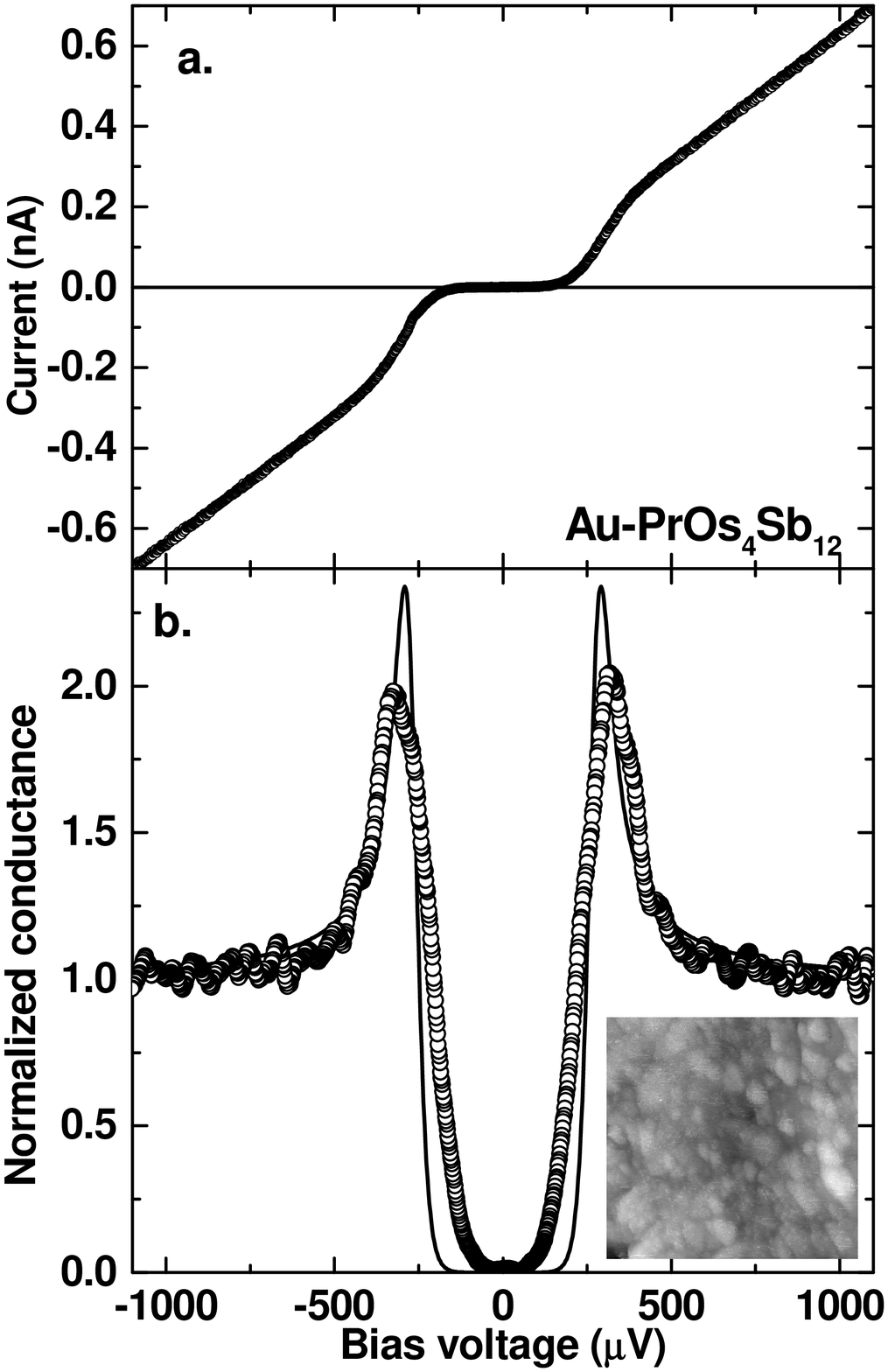}
\vskip -0.2cm\caption{Current-voltage characteristics (a) and
tunnelling conductance (b.) of PrOs$_4$Sb$_{12}$ at 0.19K. The
superconducting gap is well developed with no low energy
excitations. The line in b is the prediction from conventional
isotropic BCS s-wave theory using $\Delta=270\mu V$ and T=0.19K.
The inset in b shows a typical topographic image of the surface,
representing an area of 50nmx50nm, which shows a corrugation of
about 4 nm.} \label{fig:Fig2}
\end{figure}

Typical spectra (Fig.2) show a clean density of states with no low
energy excitations, demonstrating that the superconducting gap is
well developed over a large part of the Fermi surface. The best
fit with isotropic BCS s-wave theory is shown by the line in
Fig.2. Using T=0.19K we obtain a value for the superconducting gap
$\Delta$=270$\mu$eV, which gives $2\Delta/k_BT_c$=3.4, very close
to the BCS value 3.53. Note that differences between the
experiment and the fit
 are clearly resolved thanks to the high resolution in voltage of our experiment,
evidenced by the curves shown in Fig.1.

The tunnelling current depends on the overlap of the electron
wavefunctions of tip and sample near the surface, being therefore
a sum of the contributions from electrons coming from different
sheets of the Fermi surface having k-vectors with different
orientations. If the superconducting gap is anisotropic in a given
sheet, or it has different values in different sheets of the Fermi
surface, or both, the tunnelling spectra reflect this distribution
of values of the superconducting gap by showing more broadened
coherence peaks, which produce a deviation from isotropic BCS
s-wave behavior (Fig.2 and e.g. Ref.\cite{Hess90}).
 Note that the presence of defects, grain boundaries or other
perturbations that one can figure out to occur near the surface, can lead either to
 a decrease of the observed anisotropy
 through the mixing of the electronic wavefunctions from
different parts of the Fermi surface caused by strong inter and/or intraband electronic
 scattering\cite{Sung67,Golubov97,Martinez03,Martinez03a}, or to pair breaking.
 The former was not observed, and the
latter always leads to an increased residual density of states at the Fermi level,
 which we can indeed observe in some locations of the surface, as discussed further on.
Therefore, we believe that the deviations between the experiment and the
fit shown in Fig.2 are intrinsic to the superconducting
density of states in this compound.

Note that the conductance begins to increase at about 120$\mu$V,
and the highest point of the coherence peak is located at
325$\mu$V. As a matter of fact, the shape of the tunnelling
spectra we find is similar to the form of the spectra taken in the
much studied material
 NbSe$_2$\cite{Hess90}, where first experiments clearly showed up the presence of
a distribution of values of the superconducting gap over the Fermi surface\cite{Hess90}.
Subsequent work has identified this distribution as coming from different
 gap values in different sheets of the Fermi surface in that compound
 (multiband superconductivity)\cite{Taillefer03,Yokoya03}.
Whereas more work is clearly needed to understand the origin
of the gap distribution in PrOs$_4$Sb$_{12}$, it is noteworthy to remark that
strong changes in the mass renormalization in different
sheets has been found in the de Haas van Alphen experiments of
 Ref.\cite{Sugawara02}. These changes may also lead to the distribution of
values of the superconducting gap measured in our experiment.
This strengthens both the idea that
 the mass renormalization and superconductivity
 are of the same origin, i.e. that the quadrupolar fluctuations favor
superconducting correlations, as well as the possible multiband
character of superconductivity in this compound.

The spectra are smeared when we increase the temperature, as shown in Fig.3a,
and become flat above the bulk critical temperature (1.85K).
The maximum in the
derivative of the conductance within the quasiparticle peaks, which gives
a good estimate of the mean value of the superconducting gap \cite{Martinez03} is shown in
Fig.3b and it follows well the temperature
 dependence of the superconducting gap within
BCS s-wave theory (line in Fig.3b).
 The curve can be extrapolated to a critical temperature of 1.8K, which is
the bulk T$_c$ (1.85K) value within our experimental error (10\%).

The superconducting properties in this compound clearly differ
from the ones in magnetically mediated Ce or U heavy fermions.
Although we could not find any published STM measurements in the
tunnelling regime and in the superconducting phase in these
materials, there is compelling evidence from many different
techniques sensitive to the superconducting density of states that
a large amount of low energy excitations due to a strongly
anisotropic superconducting gap is found in most cases. For
instance, many experiments show now that the multiphase
superconductor UPt$_3$ presents a line node along the basal plane
and nodes along the c axis of its hexagonal crystalline structure,
with a superconducting gap that decreases by more than an order of
magnitude at the nodes \cite{Joynt02,SudUPt3}. In the case of
PrOs$_4$Sb$_{12}$, there are at present no data pointing towards
extended gapless regions on the Fermi surface, as the one caused
by a line of nodes, so that the superconducting gap appears to be
opened in a much larger part of the Fermi surface than in UPt$_3$.

\begin{figure}[ht]
\includegraphics[width=8cm,clip]{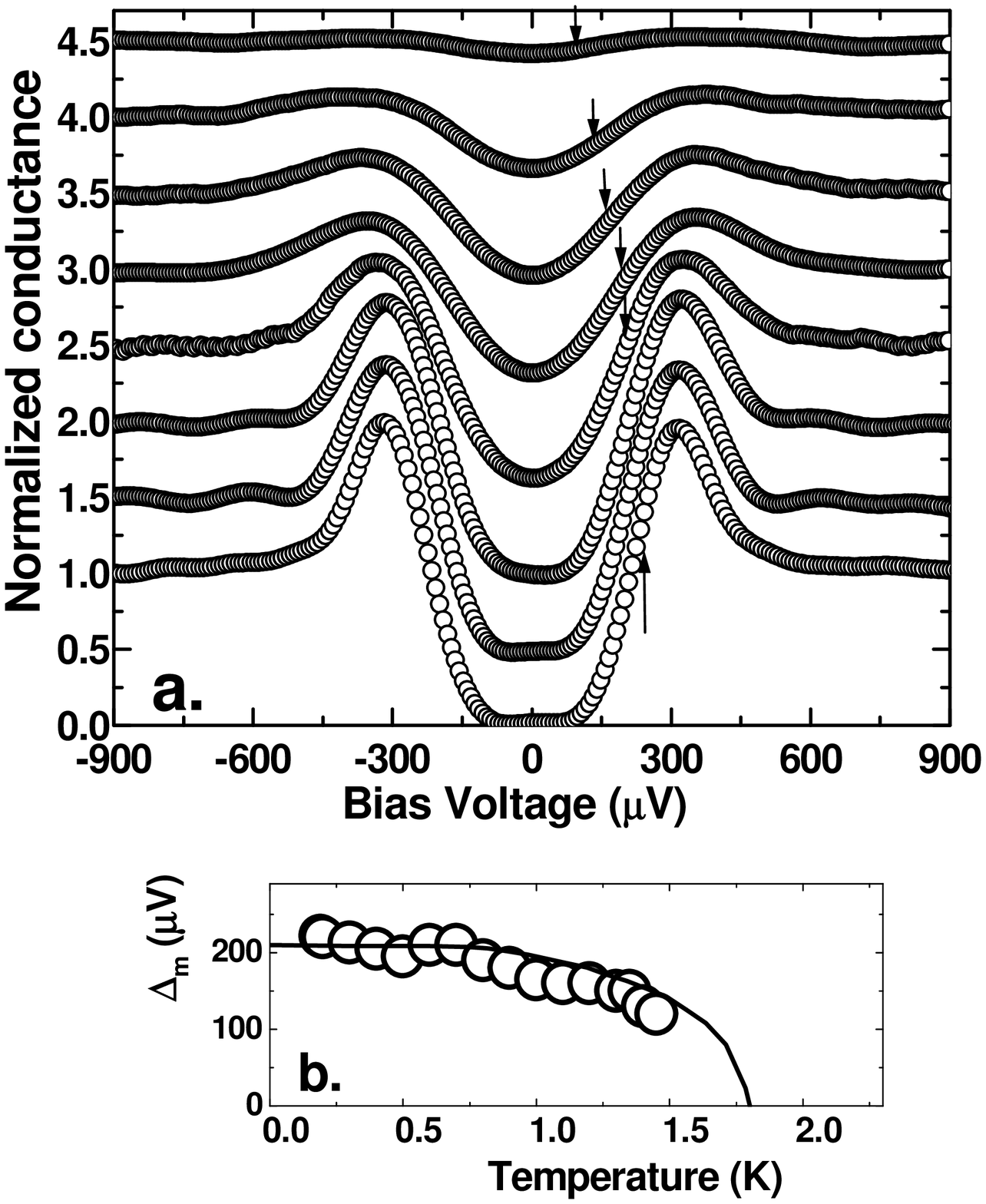}
\vskip -2.5cm \caption{in a. we show the temperature dependence of
the tunnelling conductance. The curves have been displaced by 0.5
units in the y axis for clarity. The data were taken at
0.2,0.3,0.4,0.6,0.8,1,1.2,1.4 K from bottom to top. In b., the
temperature dependence of the maximum in the derivative of the
conductance is shown (also with arrows in a.).
 This gives a good estimation of the mean
value of the superconducting gap and it fits well to
BCS theory (line). The extrapolated critical temperature coincides
 with the value found in the bulk.} \label{fig:Fig3}
\end{figure}

The observed behavior has been reproduced in different places, but
it is not found over the whole surface. In general, regions with
no residual density of states, as shown in Fig.4 a and b, have
typical sizes of several times the coherence length
($\xi_0$=12nm\cite{Bauer02}) and are surrounded by regions where a
finite density of states appears at the Fermi level, shown in
Fig.4 c and d. We can also easily find regions with much less well
defined superconducting features (not shown in the figure). When
we find a finite density of states at the Fermi level, the
superconducting features also disappear at a temperature smaller
than critical temperature of the bulk, indicating that the
physical origin for the residual density of states at the Fermi
level is some kind of strong pair breaking effect appearing near
the surface, easily detected with our technique.

\begin{figure}[ht]
\includegraphics[width=8cm,clip]{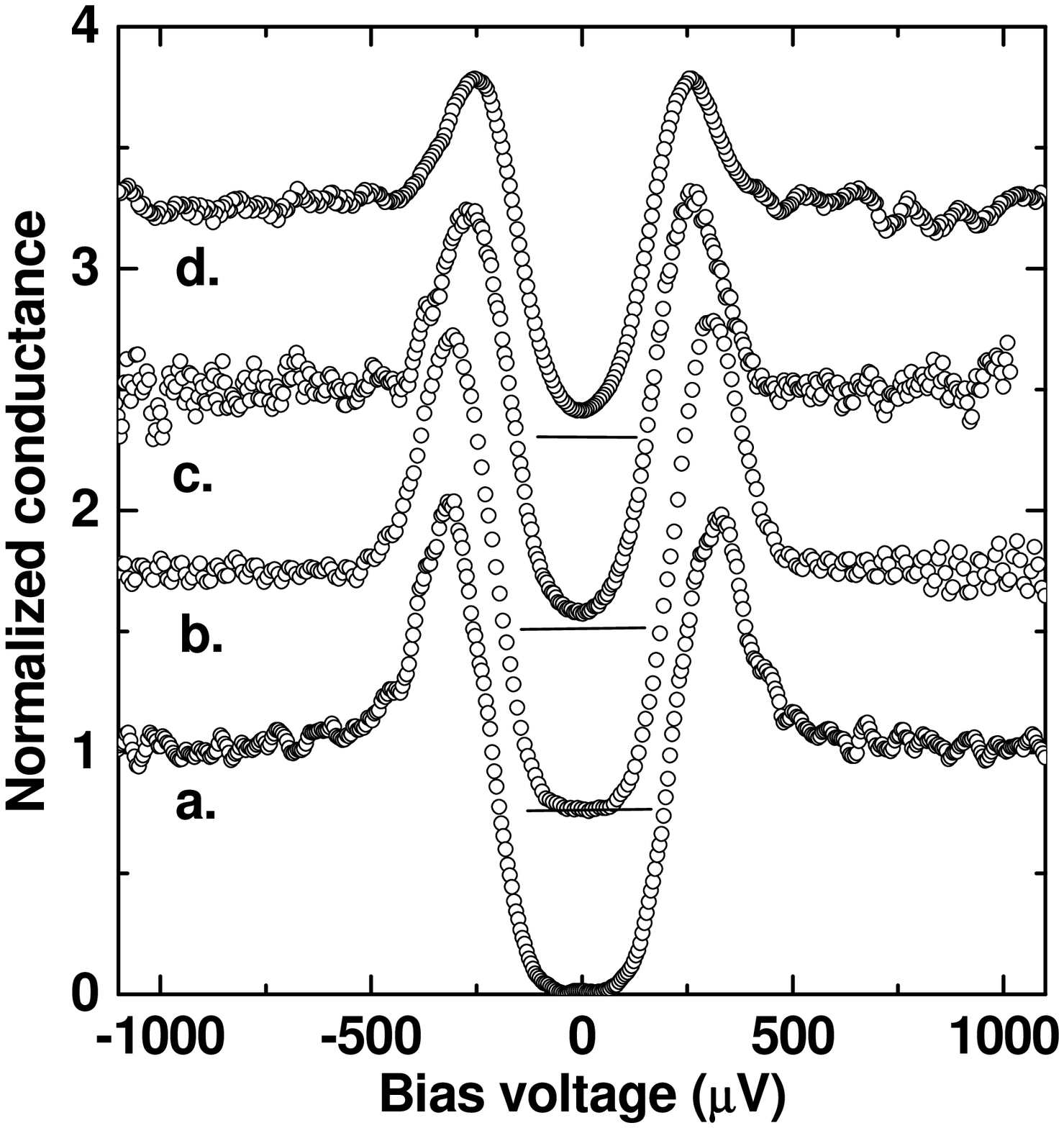}
\vskip -4cm \caption{Set of tunnelling conductance curves taken in
different positions on the surface of PrOs$_{4}$Sb$_{12}$ at
0.19K. The curves have been displaced by 0.75 units in the y axis
for clarity, and the lines show the location of zero conductance
for each curve. The finite density of states at the Fermi level
measured in some positions (c and d) shows that pair breaking
effects can appear at the surface of this compound.}
\label{fig:Fig4}
\end{figure}

In conclusion, we performed a direct measurement of the
superconducting gap through high resolution local tunnelling
spectroscopy measurements in the heavy fermion superconductor
PrOs$_4$Sb$_{12}$. Typical spectra
 demonstrate that the superconducting gap is well
 developed over a large part of the Fermi surface.
 The presence of a finite distribution
of values of the superconducting gap over the Fermi surface can be
inferred from deviations between the experiment and isotropic BCS
s-wave behavior.

We acknowledge discussions with J.P. Brison, M. Crespo,
J. Flouquet, F. Guinea, K. Izawa
 Y. Kitaoka, A. Levanyuk and J.G. Rodrigo and support from the
ESF programme VORTEX, from the MCyT (Spain; grant
MAT-2001-1281-C02-0), and from the Comunidad
Aut\'onoma de Madrid (07N/0053/2002, Spain). The Laboratorio de Bajas
Temperaturas is associated to the ICMM of the CSIC. Ames
Laboratory is operated for the U. S. Department of Energy by Iowa
State University under Contract No. W-7405-Eng-82. This work was
supported by the Director for Energy Research, Office of Basic
Energy Sciences.


\begin{thebibliography}{9}

\bibitem{Hewson}
A.D. Hewson, The Kondo problem to heavy fermions (Cambridge University Press, Cambridge,
 England 1993).

\bibitem{Aoki01}
D. Aoki et al. Nature, 413, 613-616 (2001).

\bibitem{Hasselbach89}
K. Hasselbach, L. Taillefer, J. Flouquet, Phys. Rev. Lett. {\bf
63}, p. 93 (1989); R.A. Fisher et al., Phys. Rev. Lett. {\bf 62}, 1411
(1989).

\bibitem{Yatskar96}
A. Yatskar {\it et al.}
Phys. Rev. Lett. {\bf 77}, 3637-3640 (1996)

\bibitem{Cox87}
D.L. Cox,
Phys. Rev. Lett. {\bf 59}, 1240 (1987)

\bibitem{Bauer02}
E.D. Bauer {\it et al.}, Phys. Rev. B, {\bf 65}, 100506(R), 2002.

\bibitem{Vollmer03}
R. Vollmer et al. Phys. Rev. Lett. {\bf 90}, 057001 (2003).

\bibitem{MapleLa}
M.B. Maple, Acta Physica Polonica B v. {\bf 34}, 919 (2003), cond-mat/0303370.

\bibitem{Measson}
M.A. Measson et al., to be published.

\bibitem{Aoki02}
Y. Aoki {\it et al.}, J. Phys. Soc. of Japan
{\bf 71}, 2098, (2002).

\bibitem{Joynt02}
R. Joynt, L. Taillefer, Rev. of Mod. Phys. {\bf 74}, 235-294
(2002)

\bibitem{Kotegawa03}
H. Kotegawa et al.
Phys. Rev. Lett. {\bf 90}, 027001 (2002).

\bibitem{MacLaughlin02}
D.E. MacLaughlin {\it et al.}
Phys. Rev. Lett. {\bf 89}, 157001 (2002).

\bibitem{Izawa02}
K. Izawa et al. Phys. Rev. Lett. {\bf 90}, 117001 (2003).

\bibitem{Sugawara02}
H. Sugawara {\it et al.}, Phys. Rev. B {\bf 66}, 220504 (2002).

\bibitem{Suderow02a}
H. Suderow {\it et al.}, Phys. Rev. B {\bf 65}, R100519 (2002), and references therein.

\bibitem{Wolf}
E.L. Wolf, "Principles of Electron Tunnelling Spectroscopy",
Oxford University Press (1989).

\bibitem{Rubio01}
G. Rubio-Bollinger, H. Suderow, S. Vieira, Phys. Rev. Lett. {\bf
86}, 5582 (2001).

\bibitem{Hess90}
H.F. Hess, R.B. Robinson, and J.V. Waszczak, Phys. Rev. Lett. {\bf
64}, 2711 (1990).

\bibitem{Sung67}
C.C. Sung, V.K. Wong, J. Phys. Chem. Solids, {\bf 28}, 1933
(1967).

\bibitem{Golubov97}
A. A. Golubov and I. I. Mazin, Phys. Rev. B 55, 15146-15152 (1997)

\bibitem{Martinez03}
P. Martinez-Samper {\it et al.}, Physica C {\bf 385}, 233 (2003).

\bibitem{Martinez03a}
P. Martinez-Samper {\it et al.}, Physical Review B {\bf 67},
014526 (2003).

\bibitem{Taillefer03}
E. Boaknin, et al. Phys. Rev. Lett. {\bf 90}, 117003 (2003)

\bibitem{Yokoya03}
T. Yokoya et al., Science {\bf 294}, 2518 (2001).

\bibitem{SudUPt3}
See also e.g. H. Suderow et al. Phys. Rev. Lett. {\bf 80}, 165
(1998).


\end{thebibliography}
\end{document}